# STRONG SEISMIC MOTIONS ESTIMATED FROM A ONE DIRECTION-THREE COMPONENTS ("1D-3C") APPROACH, APPLICATION TO THE CITY OF ROME, ITALY

M.P. Santisi d'Avila[1], L. Lenti[2], J.F. Semblat[3], A. Gandomzadeh[4], S. Martino[5], F. Bonilla[6]

## ABSTRACT

Strong seismic motions in soils generally lead to both a stiffness reduction and an increase of the energy dissipation in the surficial layers. In order to study such phenomena, several nonlinear constitutive models were proposed and were generally implemented for 1D soil columns. However, one of the main difficulties of complex rheologies is the large number of parameters needed to describe the model. In this sense, the multi-surface cyclic plasticity approach, developed by Iwan in 1967 but linked to Prandtl or Preisach theoretical work, is an interesting choice: the only data needed is the modulus reduction curve.

Past studies have generally implemented such models for one-directional shear wave propagation in a "1D" soil column considering one motion component only ("1C"). Conversely, this work aims at studying strong motion amplification by considering seismic wave propagation in a "1D" soil column accounting for the influence of the 3D loading path on the nonlinear behavior of each soil layer. In the "1D-3C" approach, the three components (3C) of the outcrop motion are simultaneously propagated into a horizontally layered soil for which a three-dimensional constitutive relation is used (Finite Element Method). The alluvial site considered in this study corresponds to the Tiber River Valley, close to the historical centre of Rome (Italy). The computations are performed considering the waveforms referred as the 14th October 1997 Umbria-Marche earthquake, recorded on outcropping bedrock. Time histories and stress-strain hysteretic loops are computed all along the soil column.

The octahedral stress, the strain-depth profiles and the transfer functions in acceleration (surface/outcrop spectral ratios) are estimated for the 1D-1C and the 1D-3C approaches, evidencing the influence of the three-dimensional loading path..

## INTRODUCTION

According to collected seismic records, the local site condition emerges as one of the dominant factors controlling the variation in ground motion and determination of the site-specific seismic hazard. The evidence of nonlinear soil behavior comes from experimental cyclic tests on soil samples, for different strain amplitudes (Hardin and Drnevich, 1972a, 1972b; Vucetic, 1990). The nonlinearity is observed through shear modulus reduction and increase of damping for increasing strain levels. The effect on the transfer function is a shift of the fundamental frequency toward lower frequencies, as well as an attenuation of the spectral amplitudes at high frequencies.

---

[1] University Paris-Est, Laboratoire Central des Ponts et Chaussées (LCPC), 58 Bd Lefebvre 75732 Cedex 15 Paris (France), email: santisid@lcpc.fr
[2] University Paris-Est, LCPC, Paris, e-mail: lenti@lcpc.fr
[3] University Paris-Est, LCPC, Paris, e-mail: semblat@lcpc.fr
[4] University Paris-Est, LCPC, Paris, e-mail: ali.gandomzadeh@lcpc.fr
[5] Res. Centre for Geological Risks, Univ. of Rome "Sapienza", e-mail: salvatore.martino@uniroma1.it
[6] IRSN, Fontenay-aux-Roses, e-mail: fabian.bonilla@irsn.fr

One-directional wave propagation response analyses are generally used to estimate soil surface ground motions useful for the design of structures. A complete nonlinear site response analysis requires the propagation of an earthquake record in a nonlinear medium by integrating the wave equation in the time domain and using an appropriate constitutive model. The hysteretic behavior of soils is difficult to model realistically because the yield surface can have a complex form. Some researchers adopt the theory of plasticity to describe the hysteresis of soil (Zienkiewicz et al., 1982; Chen and Baladi, 1985; Chen and Mizuno, 1990; Prevost and Popescu, 1996; Ransamooj and Alwash, 1997; Montans, 2000), others propose simplified nonlinear models (Kausel and Assimaki, 2002; Delepine et al., 2009) and other ones combine elasto-plastic constitutive equations with empirical rules (Ishihara and Towhata, 1982; Finn, 1982; Towhata and Ishihara, 1985; Iai et al., 1990a, 1990b; Kimura et al., 1993). Empirical rules that describe the loading and unloading paths in the stress-strain space are the so-called Masing rules (Masing, 1926), that reproduce quite faithfully the hysteresis observed in the laboratory (Vucetic, 1990). The main problem of these rules is that the computed stress may exceed the maximum strength of the material when an irregular load is applied (Pyke, 1979; Li and Liao, 1993). Several attempts have been done in order to control the dispersive property of these rules (Pyke, 1979; Vucetic, 1990).

One-directional models for site response analysis are proposed by several authors (Joyner and Chen, 1975; Joyner et al., 1981, Lee and Finn, 1978; Pyke, 1979; Phillips and Hashash, 2009). Furthermore, Li (1990) incorporates the three-dimensional cyclic plasticity soil model proposed by Wang et al. (1990) in a finite element procedure, in terms of effective stress, to simulate the one-directional wave propagation; however, the rheology needs between 10 to 20 parameters to characterize the soil model.

The nonlinear rheology used in this research is a multi-surface cyclic plasticity mechanism that depends on few parameters that can be obtained from simple laboratory tests (Iwan, 1967). Material properties include the dynamic shear modulus at low strain and the variation of shear modulus with shear strain. This rheology allows the material to develop large strains in the range of stable nonlinearity in undrained conditions. Because of its three-directional nature, the procedure can handle both shear wave and pressure wave simultaneously and predict not only horizontal motion but vertical settlement too.

In the present research, a finite element procedure to evaluate stratified level ground response to three-directional earthquakes is presented and the importance of the three-directional shaking problem is analyzed. The main feature of the procedure is that it solves the specific three-dimensional stress-strain problem with a one-directional approach.

## IMPLEMENTATION OF THE RHEOLOGICAL MODEL

The three components of the seismic motion are propagated into a multi-layered column of nonlinear soil from the top of the underlying elastic bedrock considering a finite element scheme. Soil is assumed to be a continuous and uniform medium of infinite horizontal extent. Soil stratification is discretized into a system of horizontal layers, parallel to the *xy* plane, using quadratic line elements with three nodes. Shear and pressure waves propagate vertically in the *z*-direction. These assumptions do not yield any strain variation in *x*- and *y*-directions.

According to a finite element modeling of a multilayer soil system assumed with an horizontal setting, the weak form of equilibrium equations, including compatibility conditions, three-dimensional nonlinear constitutive relation and the imposed boundary condition (Cook et al., 2002), is expressed in matrix form as:

$$\mathbf{M\ddot{D}} + \mathbf{C\dot{D}} + \mathbf{F}_{int} = \mathbf{F} \tag{1}$$

where $\mathbf{M}$ is the mass matrix, $\dot{\mathbf{D}}$ and $\ddot{\mathbf{D}}$ are the first and second temporal derivatives of the displacement vector $\mathbf{D}$, respectively, $\mathbf{F_{int}}$ is the vector of internal forces and $\mathbf{F}$ is the load vector. $\mathbf{C}$ is a matrix that derives from the fixed boundary condition, as explained below.

Discretizing the soil column into $n$ nodes, having three translational degrees of freedom (d.o.f.) each, yields a $3n$-dimensional displacement vector $\mathbf{D}$ composed by three blocks whose terms are the displacement of the $n$ nodes in $x$-, $y$- and $z$- direction, respectively. The assembled ($3n \times 3n$)-dimensional mass matrix $\mathbf{M}$ and the $3n$-dimensional vector of internal forces $\mathbf{F_{int}}$ respectively result from the assemblage of ($9 \times 9$)-dimensional matrices like $\mathbf{M}^e$ and vector $\mathbf{F}^e_{int}$, corresponding to element $e$, which are expressed by

$$\mathbf{M}^e = \int_H \mathbf{N}^T \rho \mathbf{N} \, dz \qquad \mathbf{F}^e_{int} = \int_H \mathbf{B}^T \boldsymbol{\sigma} \, dz \qquad (2)$$

where $H$ is the soil column height, $\rho(z)$ is the soil density and $N$ is the ($3 \times 9$)-dimensional matrix defined through the quadratic shape functions ($N_1$, $N_2$ and $N_3$) of the three-noded line element used to discretize the soil column. The terms of the ($6 \times 9$)-dimensional matrix $\mathbf{B}$ are the spatial derivatives of the shape functions, according to compatibility conditions (Cook et al., 2002). In equation (2), the stress components are terms of the 6-dimensional vector $\boldsymbol{\sigma}$.

The system of horizontal soil layers is bounded at the top by the free surface and at the bottom by the semi-infinite elastic medium which represents the seismic bedrock. The stresses normal to the free surface are imposed null and at the interface soil-bedrock the following condition, implemented by Joyner and Chen (1975) in a finite difference formulation, is applied:

$$-\mathbf{p}^T \boldsymbol{\sigma} = \mathbf{c}(\dot{\mathbf{u}} - 2\dot{\mathbf{u}}_b) \qquad (3)$$

The stresses normal to the column surface are $-\mathbf{p}^T \boldsymbol{\sigma}$ and $\mathbf{c}$ is a ($3 \times 3$) diagonal matrix whose terms are $\rho_b v_{bs}$, $\rho_b v_{bs}$ and $\rho_b v_{bp}$. The parameters $\rho_b$, $v_{bs}$ and $v_{bp}$ are the bedrock density and shear and pressure wave velocities in the bedrock, respectively. The three terms of vector $\dot{\mathbf{u}}$ are the velocities in $x$-, $y$- and $z$- direction respectively, at the interface soil-bedrock. The terms of the 3-dimensional vector $\dot{\mathbf{u}}_b$ are the input velocities, in the underlying elastic medium respectively in direction $x$, $y$ and $z$. The boundary condition (3) allows energy to be radiated back into the underlying medium. According to equation (3), the matrix $\mathbf{C}^e$ and the load vector $\mathbf{F}^e$, of each element $e$, are defined by

$$\mathbf{C}^e = \int_H \mathbf{N}^T \mathbf{c} \mathbf{N} \, dz \qquad \mathbf{F}^e = \int_H \mathbf{N}^T \mathbf{c} \mathbf{u}_b \, dz \qquad (4)$$

The finite element model and nonlinearity of the soil requires, respectively, spatial and temporal discretization, to permit the problem resolution. The relation between stress and strain increments is linearized at each time step. Accordingly, Equation (1) is expressed by

$$\mathbf{M} \Delta \ddot{\mathbf{D}}^i_k + \mathbf{C} \Delta \dot{\mathbf{D}}^i_k + \mathbf{K}^i_k \Delta \mathbf{D}^i_k = \mathbf{F} \qquad (5)$$

where the subscript $k$ indicates the time step $t_k$.

At each time step $k$, equation (5) requires an iterative solving to correct the tangent stiffness matrix $\mathbf{K}^i_k$. Starting from the stiffness matrix $\mathbf{K}^1_k = \mathbf{K}_{k-1}$, evaluated at the previous time step, the value of matrix $\mathbf{K}^i_k$ is updated at each iteration $i$ (Crisfield, 1991) and the correction process continues until the difference between two successive approximations is reduced to a fixed tolerance, according to

$$\left| \mathbf{D}_k^i - \mathbf{D}_k^{i-1} \right| < \alpha \left| \mathbf{D}_k^i \right| \tag{6}$$

where $\alpha=10^{-3}$ (Mestat, 1993, 1998). The vectors of total displacement, velocity and acceleration are respectively defined by

$$\mathbf{D}_k^i = \mathbf{D}_{k-1} + \Delta\mathbf{D}_k^i \; ; \; \dot{\mathbf{D}}_k^i = \dot{\mathbf{D}}_{k-1} + \Delta\dot{\mathbf{D}}_k^i \; ; \; \ddot{\mathbf{D}}_k^i = \ddot{\mathbf{D}}_{k-1} + \Delta\ddot{\mathbf{D}}_k^i \tag{7}$$

The stiffness matrix $\mathbf{K}_k^i$ is obtained by assembling (9×9)-dimensional matrices like the following, corresponding to element $e$:

$$k_k^{e,i} = \int_H \mathbf{B}^T \mathbf{E}_k^i \mathbf{B} \, dz \tag{8}$$

The tangent constitutive (6x6) matrix $\mathbf{E}_k^i$ is evaluated by the constitutive incremental relationship given by

$$\Delta\boldsymbol{\sigma}_k^i = \mathbf{E}_k^i \, \Delta\boldsymbol{\varepsilon}_k^i \tag{9}$$

According to Joyner (1975), the actual strain level and the strain and stress values at the previous time step allow to evaluate the tangent constitutive matrix $\mathbf{E}_k^i$ and the stress increment $\Delta\boldsymbol{\sigma}_k^i = \Delta\boldsymbol{\sigma}_k^i(\boldsymbol{\varepsilon}_k^i, \boldsymbol{\varepsilon}_{k-1}, \boldsymbol{\sigma}_{k-1})$.

The step-by-step process is solved by the Newmark algorithm, expressed as follows:

$$\begin{cases} \Delta\dot{\mathbf{D}}_k^i = \dfrac{\gamma}{\beta\Delta t}\Delta\mathbf{D}_k^i - \dfrac{\gamma}{\beta}\dot{\mathbf{D}}_{k-1} + \left(1 - \dfrac{\gamma}{2\beta}\right)\Delta t \ddot{\mathbf{D}}_{k-1} \\ \Delta\ddot{\mathbf{D}}_k^i = \dfrac{1}{\beta\Delta t^2}\Delta\mathbf{D}_k^i - \dfrac{1}{\beta\Delta t}\dot{\mathbf{D}}_{k-1} - \dfrac{1}{2\beta}\ddot{\mathbf{D}}_{k-1} \end{cases} \tag{10}$$

Parameters $\beta=0.3025$ and $\gamma=0.6$ guarantee a conditional numerical stability of the temporal integration scheme with numerical damping (Hughes, 1987). Further research is required to investigate the influence of numerical damping on the solution and optimize the couple of chosen parameters. Equations (5) and (10) yield

$$\bar{\mathbf{K}}_k^i \, \Delta\mathbf{D}_k^i = \Delta\mathbf{F}_k + \mathbf{A}_{k-1} \tag{11}$$

where the modified stiffness matrix is defined as

$$\bar{\mathbf{K}}_k^i = \frac{1}{\beta\Delta t^2}\mathbf{M} + \frac{\gamma}{\beta\Delta t}\mathbf{C} + \mathbf{K}_k^i \tag{12}$$

and $\mathbf{A}_{k-1}$ is a vector depending to the response in previous time step, given by

$$\mathbf{A}_{k-1} = \left[\frac{1}{\beta\Delta t}\mathbf{M} + \frac{\gamma}{\beta}\mathbf{C}\right]\dot{\mathbf{D}}_{k-1} + \left[\frac{1}{2\beta}\mathbf{M} + \left(\frac{\gamma}{2\beta} - 1\right)\Delta t \mathbf{C}\right]\ddot{\mathbf{D}}_{k-1} \tag{13}$$

After evaluating the displacement increment $\Delta \mathbf{D}_k^i$ by equation (11), using the tangent stiffness matrix corresponding to the previous time step, velocity and acceleration increments can be calculated by equation (10) and the total parameters are obtained according to (7). The strain increment $\Delta \boldsymbol{\varepsilon}_k^i$ is deduced by the displacement increment $\Delta \mathbf{D}_k^i$ and the stress increment $\Delta \boldsymbol{\sigma}_k^i$ and tangent constitutive matrix $\mathbf{E}_k^i$ are obtained by constitutive relationship (14). The modified stiffness matrix $\overline{\mathbf{K}}_k^i$ is calculated and the process restarts until condition (6) is verified. Afterwards, the next time step is analyzed.

## FEATURES OF THE CONSTITUTIVE MODEL

The nonlinear soil behavior in a three-dimensional stress state is properly modeled by the constitutive model suggested by Iwan (1967) and applied by Joyner and Chen (1975) and Joyner (1975) in a finite difference formulation. This model takes into account the nonlinear hysteretic behavior of soils, using an elasto-plastic approach with hardening based on the definition of a series of nested yield surfaces. The main feature of this rheological model is its flexibility for incorporating laboratory results on the dynamic behavior of soils. The only necessary input data, to identify soil properties in the applied constitutive model, are the mass density $\rho$, shear and pressure wave velocities in the medium, $v_s$ and $v_p$ respectively, and the normalized shear modulus decay curve $G/G_0$ versus shear strain $\gamma$. $G_0 = \rho v_s^2$ is the elastic shear modulus, measured at the elastic behavior range limit $\gamma \cong 10^{-6}$ (Fahey, 1992).

The homogeneous cyclic soil response to one-component signal is evaluated to observe nonlinear effects. A sinusoidal shear strain input with increasing amplitude (Figure 1a) is applied in $x$-direction and the cyclic strain-stress behavior is displayed in Figure 1b. We assume that the behavior of soil can be adequately represented by a hyperbolic stress-strain curve. According to Hardin and Drnevich (1970), this assumption yields a normalized shear modulus decay curve expressed as (Figure 1c)

$$G/G_0 = 1/\left(1+|\gamma/\gamma_r|\right) \tag{15}$$

where $\gamma_r$ is a reference strain provided by test data for a shear modulus reduction of $50\%$. The soil properties are $\rho = 1780 \, \text{kg/m}^3$, $v_p = 560 \, \text{m/s}$, $v_s = 300 \, \text{m/s}$ and $\gamma_r = 0.1\%$ and the input frequency 3 Hz.

Applying cyclic shear strains with amplitude greater than the elastic behavior range limit gives open loops of shear stress versus shear strain, exhibiting some hysteresis. As strain amplitude is increased, the shear modulus reduces. This reduction is represented by a normalized shear modulus decay curve (Figure 1c).

The homogeneous cyclic soil response to a three-component seismic signal is evaluated to analyze nonlinear effect variation under triaxial stress state. The soil response to one-component signal is compared with the case of two equal shear strain components applied in $x$- and $y$-directions and a normal strain component reduced by a factor of 10 applied in $z$-direction (Figure 1d, e, f). The material strength is lower under triaxial loading than for simple shear loading. Figure 1c shows the normalized shear modulus decay curve used as input and the normalized tangent shear modulus decay and damping in the two cases of one- and three-component shear strain input. The material damping ratio $D$ represents the energy dissipated by the soil. It is evaluated as

$$D = W_D/(4\pi W_S) \tag{16}$$

where $W_D$ is the energy dissipated in one cycle of loading and $W_S$ is the maximum strain energy stored during the cycle.

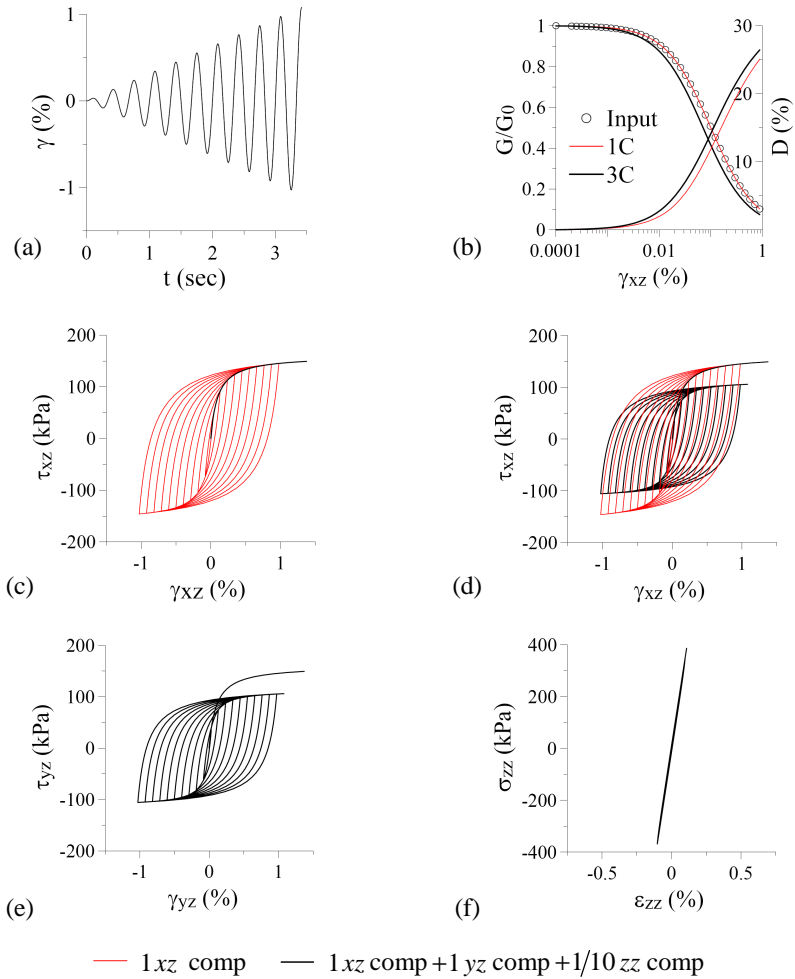

**Figure 1. Three-component effect in material behavior: a) Increasing amplitude sinusoidal strain input; b) Normalized shear modulus decay and damping computed in the two cases of one- and three-component shear strain input; c) Hysteretic response to one-component input; d, e, f) Comparison between the hysteretic response in *x*, *y* and *z* direction, respectively, in the two cases of one- and three-component input.**

The shear modulus decreases and the dissipation increases, for increasing strain amplitude, due to nonlinear effects also when only one component is applied. From one to three components, for a given maximum strain amplitude, the shear modulus decreases and the dissipation increases.

The method does not depend on the hyperbolic initial loading curve. Different relationships satisfying Masing criterion can be represented by a model of the Iwan type and a purely empirical stress-strain curve derived from laboratory measurements can also be directly used. Stress-strain curves of real soils are not usually truly hyperbolic, thus it is more convenient to directly apply the real shear modulus decay curve obtained by test data.

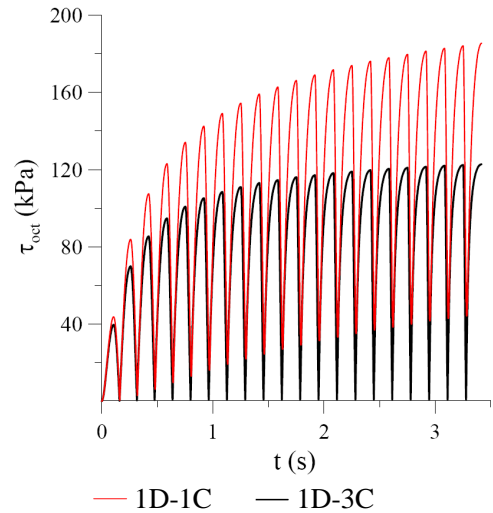

**Figure 2. Time history of the octahedral shear stress for the case of three combined 1D-1C seismic response analyses, in x-, y- and z-direction respectively, and for the 1D-3C case.**

## DESCRIPTION OF THE SITE AND INPUT

The stratigraphic setting of a soil column in the Tiber Valley of Rome (Italy) is used to analyze the seismic wave propagation in the cases of one- and three-component input and to compare the obtained results with those provided by public nonlinear codes, in the case of one-component input. The spatial description of stratigraphy and lithology of the alluvial deposits in the Tiber Valley of Rome, is described by Bonilla et al. (2010). The soil column that is modeled in this study consists of five layers on a seismic bedrock, whose depth below the ground level $z$ and physical properties, as density $\rho$, shear velocity in the medium $v_s$, and the elastic shear modulus $G_0$, are reported in Table 1 according to Bozzano et al. (2008). The pressure wave velocity in the medium $v_p$ is deduced by imposing a Poisson's ratio of 0.3. Site and laboratory testing of the Tiber alluvial deposits (Bozzano et al., 2000, 2008) shows a significant stiffness contrast between the sand layers (lithotypes R, A, B and D), silty-clayey alluvia (lithotype C) and sandy gravels of the bedrock. The basal gravels are considered as the local seismic bedrock, being characterized by a S-wave velocity $v_s > 700 \, \text{m/s}$.

The dynamic mechanical properties of the Tiber alluvial deposits come from laboratory data obtained by resonant column (RC) or from cyclic torsional shear tests. The normalized shear modulus decay curves employed in this work are shown in Figure 3.

**Table 1. Stratigraphic and geotechnical properties of the soil column (Tiber valley, Rome)**

| Layer | z | $\rho$ | $v_s$ | $G_0$ |
|---|---|---|---|---|
| | m b. g. l. | kg/m$^3$ | m/s | MPa |
| R | 0 - 2.5 | 1830 | 220 | 89 |
| A | 2.5 - 12.5 | 1875 | 239 | 107 |
| B | 12.5 - 32.5 | 1865 | 417 | 324 |
| C | 32.5 - 52.5 | 1865 | 212 | 84 |
| D | 52.5 - 57.5 | 1957 | 417 | 340 |
| Rock | > 57.5 | 2141 | 713 | 1088 |

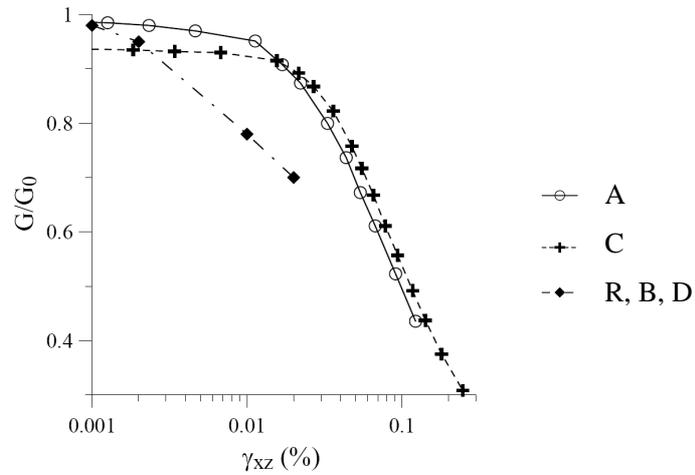

**Figure 3. Normalized experimental shear modulus decay curves
for the soils present in the analyzed column.**

The acceleration time histories in Figure 4 are the three components of the seismic event referred as the 14[th] October 1997 Umbria-Marche earthquake, characterized by a horizontal PGA of 0.3 g and local magnitude of 5.4 ML. These waveforms are provided by the ENEA ground motion database, recorded on outcropping bedrock by CODISMA digital accelerometer devices (Lenti et al., 2009). Accelerations are recorded at Cerreto di Spoleto (Perugia), about 25 km far from the earthquake epicenter. The three acceleration time histories in Figure 4 have been filtered using a cut-off frequency of 20 Hz. These acceleration signals are halved to take into account the free surface effect and integrated, to obtain the corresponding input data in terms of vertically incident velocities, before being forced at the base of the horizontal multilayer soil model.

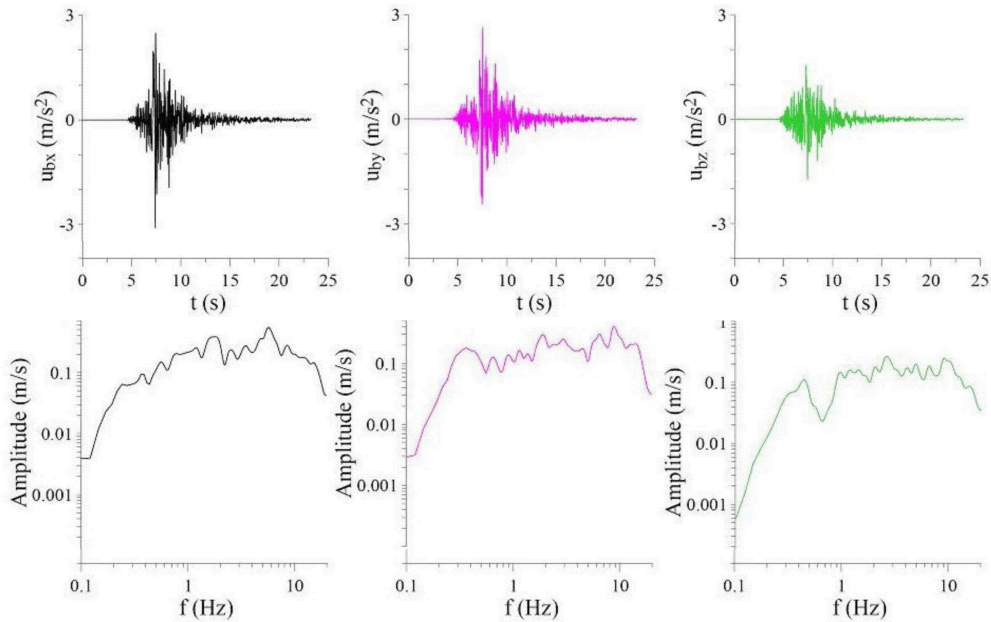

**Figure 4. Recorded acceleration time history and corresponding Fourier Transform
in West-East, North-South and vertical directions, respectively**

## ANALYSIS OF THE LOCAL 3C SEISMIC RESPONSE

The effect of simultaneously propagating the three components of the input signal in a system of horizontal soil layers is then studied and compared to the cases of one-component input. The three-component incident motion displayed in Figure 4, registered on outcropping rock, is propagated in the vertical profile of Tiber Valley, in Rome (Bonilla et al., 2010), described in section 4. The finite element procedure described in section 2 is applied for a one-directional wave propagation problem for local seismic response analysis.

Acceleration time history at the free surface, stress time history and hysteresis loop, in the middle of the C clay layer, are compared, in $x$ - direction, for the case of one- and three-component input (Figure 5a, c, d). Strain and stress profiles are shown in Figure 5b.

The maximum octahedral shear strain and stress distribution, obtained along the vertical soil profile, is shown in Figure 6 for the two cases of three combined 1D-1C analyses in the three directions of input motion and of 1D-3C analysis. The low strain level that is attained in the analyzed soil column does not allow discerning the different achieved soil strength. Parametric studies for different applications are necessary to corroborate the model for higher deformation values.

The theoretical transfer function is a technique that is frequently used for site response estimation. This approach considers the ratio between the estimated acceleration spectrum at a site of interest and the recorded acceleration spectrum at a reference site, which is usually a nearby rock site. According to this criterion, the local site responses along the studied soil profile are compared in terms of horizontal acceleration amplification functions.

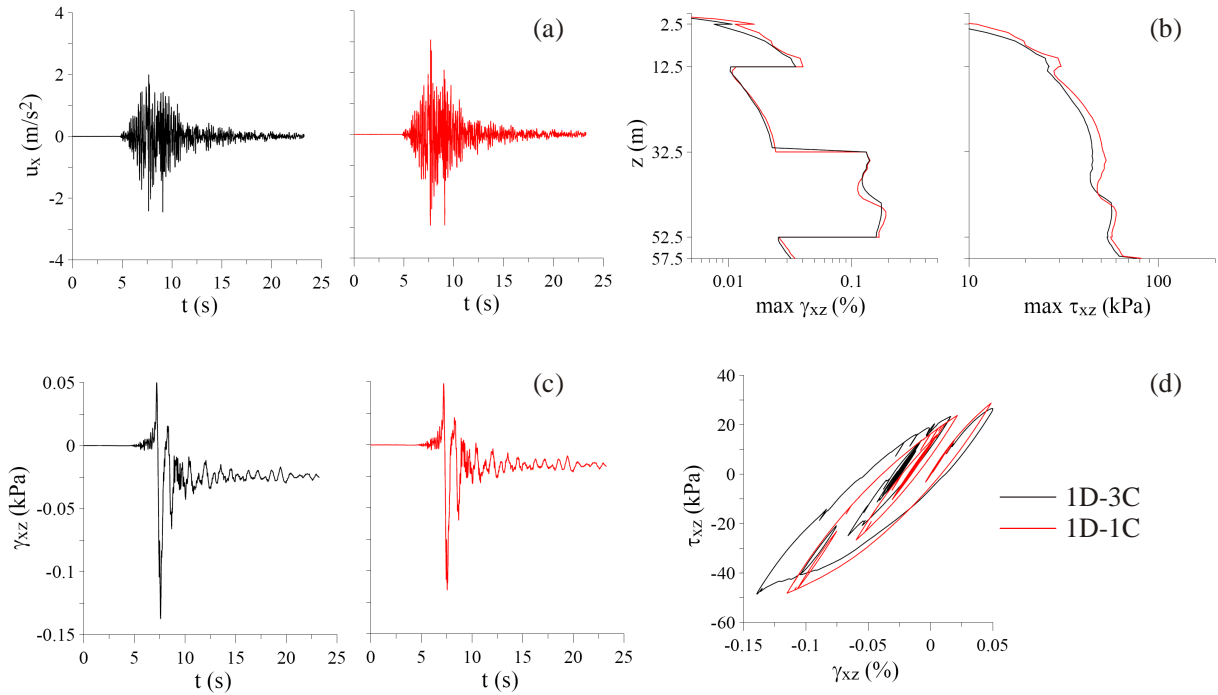

**Figure 5. Comparison between one- and three-component seismic response analysis in x-direction: a) Acceleration at the free surface; b) Maximum shear strain and shear stress profiles; c, d) Time history of the shear strain and hysteretic loop, respectively, computed at 42.5 m depth.**

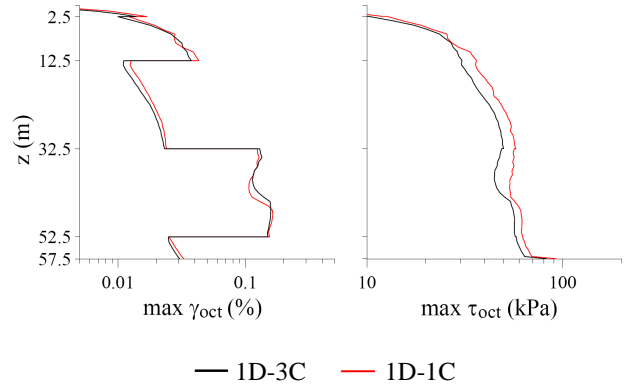

**Figure 6. Maximum octahedral shear strain and stress profiles for the case of three combined 1D-1C seismic response analyses, in x-, y- and -direction respectively, and for the case 1D-3C**

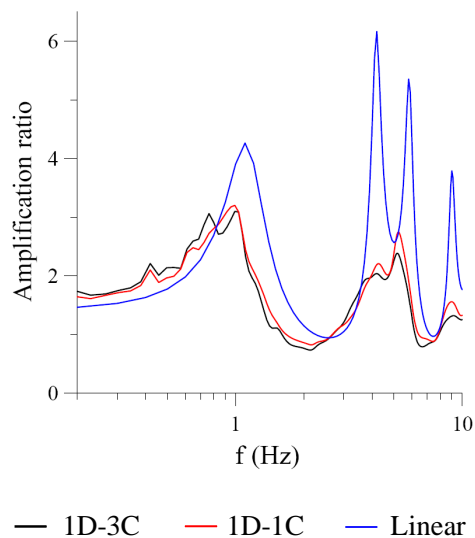

**Figure 7. Norm of horizontal acceleration transfer functions for a linear behavior, for the case of two combined 1D-1C seismic response analyses in x- and y-direction, and for the 1D-3C case**

The norm of the transfer functions of two horizontal accelerations, from the 3C results, is compared to the 1C transfer function for the linear and nonlinear computations (Figure 7). Nonlinearity effects produce a shift of the fundamental frequency toward lower frequencies, as well as an attenuation of the spectral amplitudes at higher frequencies. The effect on the transfer function produced by a three-component seismic input can not be defined using one case only, for this reason further research is necessary to compare the results obtained in different cases and to implement this nonlinear approach in two- and three-directional models.

## CONCLUSIONS

In this paper, a mechanical model is proposed to analyze the 1D-3C seismic response of soil profiles. A finite element modeling of a horizontal multilayered soil is implemented, by adopting a three-dimensional constitutive relation that needs few parameters to characterize the hysteretic behavior of the soil.

The proposed method provides a promising solution for local seismic response evaluation and site effect analysis, useful for structural design. This work is a natural extension of the public nonlinear codes such as NERA. Parametric studies and comparative analysis with experimental data are still necessary to calibrate the solution and to evidence the three-component effects in the 1D-3C approach. Efficient finite element formulation of the proposed mechanical model for two- and three-directional cases motivates further developments.